# ‘OUMUAMUA AND SCOUT ET PROBES

**JOHN GERTZ** Foundation for Investing in Research on SETI Science and Technology (FIRSST), website: www.firsst.org. c/o Zorro Productions, 125 University Avenue, Suite 101, Berkeley, CA, 94710 USA.

**Email** jgertz@firsst.org

Numerous advantages may accrue to ET from the use of probes for the conveyance of information relative to the alternative of beaming information electromagnetically from its home star system. However, probes would benefit from a trans-galactic communications network to transmit information back to their progenitor civilization(s) and throughout the Galaxy. A model for such trans-galactic communications has previously been proposed, comprised of probes linked together by relay nodes. The current paper refines that model, specifying two types of probes or nodes, both part of a communications lattice. The first type of cellular probe (CP) is statically associated with individual stars, and the second type is a scout CP that traverses the galaxy surveying successive star systems rather than permanently residing within any one of them. CPs would communicate with adjacent CPs and, ultimately, through the network of CPs, with their progenitor civilization(s). The question of whether ‘Oumuamua is such a flyby scout CP is considered. Search strategies for CPs are suggested.

**Keywords:** SETI, ET, Nodes, Probes, ‘Oumuamua

## 1 INTRODUCTION

There are strong arguments that ET might prefer to send information-laden probes to our Solar System rather than communicate by beaming radio or laser messages from its home star system [1, 2, 3, 4, 5, 6, 7, 8, 9]. For example:

- Probes might conduct surveys of a solar system even in the absence of finding a communicative civilization.

- In the event that ET’s probe does encounter a technological civilization, it can enter into an immediate dialogue unencumbered by hundreds or thousands of years of delay engendered by the speed of light.

- Probes offer security for the sender as they need not reveal their progenitor’s coordinates.

- The cost per bit of information conveyed is likely to be much cheaper than that beamed by radio or laser beacon. The envisioned comparison is between a radio or laser signal directed at Earth continuously for millions or billions of years versus the cost of sending a single light weight probe at low speed from a low gravity launch site.

- The transmitting civilization need only launch its probe once, rather than be dedicated to broadcasting over potentially millions or billions of years.

- A remote civilization will face severe, if not insurmountable problems in knowing how to communicate to an unstudied recipient civilization in a fashion that will be understood. A probe can first study its target civilization at close range and thereafter communicate in a fashion that can be understood; perhaps even in the local language (in the case of Earth, English or some other terrestrial language).

- Were ET to transmit via beacon it would require one gargantuan receiver for each star it had targeted (perhaps numbering in the millions). These would have to be dedicated receivers because ET would have no idea when a response will be received. Each receiver would have to be gargantuan because it must be capable of deciphering an Arecibo class signal, since the civilization they will hear from is likely to be nascent and not capable of much more. ET would then face the same challenge as Earth in correctly decoding any message.

- The sending civilization need not itself persist once its probe(s) is launched and thus co-exist in time with the recipient civilization, thus solving for the problem of L in the Drake Equation.

Probes, however, as classically postulated by Bracewell [1] and others, suffer from one possible challenge: communication would likely be one-way only. Earth would receive as much information as the probe wished to impart, but the probe’s progenitor civilization would learn nothing about us unless it had equipped its probe with a massive radio or laser transmitter.

To overcome this challenge, a second type of probe, a node, could function as a signal relay station. Nodes might be resident in orbit around every star, or around stars spaced as far from one another as ET’s cost/benefit analysis might determine to be most efficient. These spacecraft would not just contain the information with which they were launched, but could acquire information indefinitely from adjacent spacecraft, and through those, ultimately acquire input from an entire communication network spread throughout the Galaxy. As such, they would act as galactic libraries, constantly growing in content, irrespective of whether the total number of extant contributing civilizations increased or decreased over time. These spacecraft might function both as probes, collecting information about the solar systems that they pass through, and as nodes, passing information along to similar nearby spacecraft. Like cell phone cameras and voice recorders, they could collect data, as well as both receive data and transmit data, much as smartphones pass along data through towers and the Internet. The nomenclature of "probes" and "nodes" might then be replaced by a new term, cellular probe (CP).

## 2 NOMADIC CELLULAR SCOUT PROBES

Land based AI or land based biological beings would presumably construct and launch CPs.

The CP hypothesis as previously postulated was more or less static, with CPs associated with individual stars or small groupings of stars, such that the distance between them is approximately equal to the maximum distance allowable for reliable communication. To this structure is now added fleets or waves of scout CPs that would traverse the galaxy searching for nascent technological civilizations. CPs might fly in squadrons or lines or be spaced like pearls on a string, so that they might pass along data from one scout to the next until it arrived at a static CP. From there, data would traverse the entire galactic network from one static CP to the next and back to the land bases from whence they were launched.

In this manner, all CPs, whether of the scout or static variety, would be in communication with all other CPs, and eventually with the civilization(s) to which they belong. Each scout CP might pass close enough to each star in its path to bend it into a new trajectory toward the next star, with trajectory bending presumably being aided by a solar sail or some other propulsion system. A single scout CP could fly by an arbitrarily large number of stars, sequentially harvesting energy and receiving a gravity slingshot assist with each stellar bypass. As each scout CP caromed from one star to the next, it could gain speed and direction from the velocity vector of each star. A neatly organized wave of scout CPs might soon become disorganized as its members ricocheted off successive stars. While CPs may generally remain packed tightly together in order to facilitate communication, any scout CP that travelled too far afield from its group might potentially upload its collected data to a static CP the next time it came within communication range. CPs could ricochet among the stars indefinitely. However, they would face an ultimate speed limit of 537 km/s, the galactic escape velocity, beyond which they would be ejected from the galaxy. If CPs are easily replaceable, perhaps they might be sacrificed in an effort to sample more star systems within a given unit of time.

After one or a few passes, in order to optimize efficiency over time, many or most star systems might be dropped from the master destination list as uninteresting, for example, because they lack biological (or at least habitable) planets or moons. The Sun would presumably have remained on the refined flyby list because an earlier wave of CPs would have determined that it harbored a planet with a biosphere, or at least that it has large amounts of liquid water on its surface as well as plate tectonics, which would be obvious from multiple flybys over deep time.

Consequently, the ratio of CPs to stars required in order to adequately cover any desired region of the Milky Way is much less than 1:1. Waves of scouts need not be sent very often. Perhaps Earth is visited only once in a 100,000 years, which would seem to be a more than adequate time interval prior to the determination that at least one Earth species had become technologically active. After all, save for the last ~10,000 years, Earth's biological changes as might be recorded by a scouting CP occurred at such a slow pace that sampling at 100,000 year intervals would show only gradual changes, much like the individual frames of a movie. It is possible that visits were paced in accordance with life markers. For example, flybys might have occurred only once in 100 million years until the detection of significant atmospheric oxygen, and then increased to once in one million years thereafter until the passage of enough time, as predicted by ET's models of evolution, for the emergence of mega fauna. Thereafter, flyby frequency might have been increased significantly or a permanent static CP installed.

The point is that waves of scouting CPs, each close enough to another to facilitate communication, may make better economic sense than the classic probe model of one civilization sending a message in a bottle to another civilization. Were ET to employ scout probes and it were to encounter a nascent civilization emitting electromagnetic (EM) transmissions, the scout would require only one modest sized receiver/transmitter pointed at a nearby relay node. Translations and formatting would take place by this local discovery probe and transmitted via a system of relay nodes and in a format that ET will understand [10, 11].

Whatever their number, no one alien civilization may be able to maintain such a vast lattice of static and scouting CPs. However, it may well be within the capacity of a galactic club of civilizations acting in concert. Earth may one day be invited to join that club. Indeed, playing its role in the maintenance of the galactic communications architecture may be Earth's price of admission. In addition to normal maintenance, Earth might house large server farms of data, be a construction and launch site for new CPs, especially if new scouts are needed to replace CPs that are lost from the galaxy because they are designed to exceed its 537 km./sec. escape velocity.

Once a technological civilization is located, scouts might be superseded by classical probes left to orbit the local star or its technologically active planet(s) or moon(s).

A permanent CP might be deployed in our Solar System in at least one of three ways:

- A passing scout CP might deploy a sub-CP. As a reasonable guess, a flyby CP would have detected Earth's technological EM emissions from a distance of at least one LY, and possibly much further. It could then release a sub-CP to use its own propulsion system to decelerate. This would apply to future CPs, but not necessarily to 'Oumuamua (considered below) because the earliest it would have been able to detect Earth's EM would have been about 1940, at which time it was <.01LY from Earth.

- The first scout CP to detect Earth's EM might itself decelerate into a Solar or Earth orbit. The feasibility for this entirely depends on the distance at which the EM is detected, the velocity of the CP at the time of detection, and the ability of the CP's propulsion system to decelerate from that distance within the time necessary to do so.

- The first scout CP to detect Earth's EM might pass the detection along the communications lattice such that another scout CP can make the necessary course and velocity corrections to eventually become a resident CP within our Solar System. For example, were 'Oumuamua a scout CP, it could pass information about Earth along to the next CP scheduled to pass Earth some thousands or tens of thousands of years hence, or to some other nearer scout that might course correct. That CP would have ample time to decelerate. Alternatively, if the discovering CP cannot decelerate in time it might instead ricochet off a nearby star in such a fashion as to return to our Solar System, decelerating en route.

## 3 IS 'OUMUAMUA AN ET SCOUT CELLULAR PROBE?

'Oumuamua (officially known as 1I/2017 U1) is the first object observed within our Solar System that has been conclusively determined to have originated from elsewhere in the Galaxy. Although it entered our Solar System about 40,000 years ago, it was only first observed on October 19, 2017, after it had already passed near the Sun on its way out of the Solar System [12].

Abraham Loeb triggered international headlines in late 2018 when he made the case that 'Oumuamua may be of artificial construction [13]. Specifically, he and a colleague postulated that the available, but scanty, evidence supports the thesis that 'Oumuamua is a solar sail detached from its mother ship—in effect, alien space junk. The reasoning behind deeming this to be "junk" is that 'Oumuamua is erratically tumbling, and it is hard to imagine how a tumbling solar sail could be of any functionality. Loeb's evidence for its artificial origin includes its shape, which might be a flattened oval, and hence, to his mind, a solar sail; its unusual shininess; and the small (<0.1%) unexplained deviation in its trajectory as it passed the Sun relative to what would have been expected due to the effects of gravity alone.

### 3.1 'Oumuamua Does Not So Neatly Fit Loeb's Space Junk Model

Loeb's determination of this to be space junk flies in the face of another of his evidentiary supports, namely, that if 'Oumuamua were ejecta from a forming star system, there would have to be $10^{15}$ such pieces of ejecta per star system for one to be expected to meander through our Solar System at this time. However, what is true about ejecta would be equally true for unguided space junk. There would have to be $10^{15}$ such pieces of space junk per star for one such piece to coincidentally intersect our inner Solar System at this time. New Horizons has recently discovered that Kuiper object 2014 MU69 (Ultima Thule) is shaped like a flattened pancake, undercutting Loeb's position that 'Oumuamua's possibly similar shape may not be natural.

### 3.2 'Oumuamua Is Not Likely to Be a Functioning Spacecraft

Even discounting Oumuamua's tumbling, it still does not fit any model for what might be expected from a functioning alien probe surveilling our Solar System.

- Attempts have been made to listen for narrowband radio messages from 'Oumuamua by the SETI Institute's Allen Telescope Array (ATA) and Breakthrough Listen's Green Bank Telescope, and none have been detected (Green Bank's detection threshold was ~1000 times greater than that of the ATA) [15].

- A surveillance probe might be expected to enter the Solar System along the plane of the ecliptic for its best view of the planets. Oumuamua's entry and exit inclinations were far from the ecliptic.

- An alien probe on a flyby mission through our Solar System might be expected to be coming from some other nearby star and to be headed toward some nearby star. If, for example, 'Oumuamua were coming from a close encounter with Alpha Centauri (4.3 light years), and headed for another close encounter with Barnard's star (6.0 light years), it might be interpreted as almost irrefutable evidence in favor of a first detection of ET. In contrast, 'Oumuamua seems to have come from nowhere in particular, and is now headed toward nowhere in particular. Extrapolations of its trajectory in both directions show that it has not passed and will not pass within the inner solar system of any other star in at least the past or future one million years, and probably much longer [15].

- An alien CP might be expected to slow down as it entered the inner Solar System in order to better surveil and communicate, and to speed up as it left in order to get to the next destination as quickly as possible. 'Oumuamua has done the opposite, fully obeying the laws of gravity (apart from the, as yet, not fully explained small deviation of <0.1%).

- Moreover, its interstellar cruising speed relative to the Sun is just 26.33 km/s, not much greater than the New Horizons spacecraft now headed toward interstellar space at 16.26 km/s. At 26.33 km/s, it would take almost 50,000 years to arrive at Alpha Centauri, were it headed there, which it is not. Surely, an advanced civilization going to all the trouble of sending a scout CP can do better than that. There should be no reason that CPs could not attain the maximum speed of 537 km/s if they are not allowed to escape the galaxy, and much more if they are expendable.

### 3.3 Some Speculations that Favor 'Oumuamua as an ET Probe

- Although 'Oumuamua did not enter our Solar System along the ecliptic and therefore missed an opportunity to closely surveil most planets, it did pass within 0.16 AU, or about 15 million miles, of Earth, the planet of probable most interest to aliens.

- Although 'Oumuamua is traveling at a relatively lazy speed for practical interstellar travel and headed to no known destination, it is still possible that it will use an onboard propulsion system to speed up and change course at some point in time.

- Although 'Oumuamua is not coming from nor headed to any nearby stars, the model developed in Section 2 above allows for the possibility that CPs only fly by stars of intense interest, presumably because they contain biological planets. It may well be that there are no other such planets in our neighborhood.

- Although 'Oumuamua is tumbling, suggesting that if it is a solar sail it is non-functioning one, it may not have randomly entered our Solar System. It is possible that it was intentionally or unintentionally jettisoned after its associated probe was already inside or headed directly toward our Solar System [16].

## 4 SEARCH STRATEGIES

Although the preponderance of evidence does not support the hypothesis that 'Oumuamua is a CP of alien derivation, further data collection is warranted:

- It seems only prudent to continue to monitor 'Oumuamua for radio, IR or laser transmissions.

- Ideally, 'Oumuamua should be observed for accelerations or a change in trajectory not explained by gravity. Unfortunately, it is already beneath the detection threshold of Hubble and all terrestrial telescopes unless very large amounts of telescope time would be devoted to the task of further observing such a dim object.

- Direct observations were made by various observatories of the departing Oumuamua. A small (<0.1%) deviation from the course predicted by gravity alone was observed. Although this can most probably be explained by outgassing, no actual outgassing was observed [17]. Available databases should be examined for the incoming 'Oumuamua to determine whether there were any other deviations in course or velocity unexplained by gravity or outgassing.

- Were 'Oumuamua a transiting alien spaceship (assuming that it is actually functioning rather than inert space junk), it might have deposited a smaller probe into a tight solar orbit once it had detected Earth's technological EM, such as originates from terrestrial television broadcasts and radar. That probe might be left behind to further surveil Earth, decode its EM and eventually communicate with it. 'Oumuamua might be akin to a Pony Express rider, dropping off the mail as it rode by. Therefore, we should monitor the area immediately around the Sun for such a left behind probe. A tight solar orbit would allow it to harvest energy. I have argued elsewhere that tight solar orbits might be preferred so that adjacent CPs know exactly where to aim narrow beamed lasers [11]. Had 'Oumuamua left behind a CP, that CP might even now have begun to study our languages and number system from Sesame Street, our math from Khan Academy, and the rest from YouTube and Wikipedia. Observations of the space around the Sun are warranted in any event, since even if 'Oumuamua is not itself an alien spaceship or CP, an actual alien CP may have been deposited at any time in the past.

- 'Oumuamua might be a transiting CP, while a permanent CP might reside on an asteroid, the Moon, in a tight orbit around the Sun or elsewhere within our Solar System. It might even be in Earth orbit and have been misclassified as space junk. Transiting or permanent CPs within our Solar System presumably have detected our EM and know we are here. It might therefore seem reasonable to target local objects with METI transmissions [18]. It is possible, as METI proponents have suggested, that ET protocols include a proscription against transmitting to Earth unless and until Earth intentionally transmits first. Perhaps this applies equally to local CPs as it does to a remote ET. However, METI transmissions aimed at targets within our Solar System should not inadvertently illuminate background stars. Therefore, they should transmit at a power that is no higher than Earth's omni-directional EM leakage, or if at higher power, never while the target is traversing the plane of the Milky Way or occulting a nearby star. Local METI might be encouraged on three additional grounds: (a) It gives METIists practice in message construction and transmission; (b) there is at least as good a chance that ET is sending probes rather than beaming transmissions from their home planets so their efforts would not be assumed to be wasted; and (c) receipt of a return message would be methodologically feasible since when transmitting to the stars one must reserve a telescope for the return message at a future time that is equal to at least twice the time for a transmission to arrive at each target star (which is completely impractical). If anti-METIists (this author included) were amenable to restricted local METI this might form the basis of a grand compromise between the warring camps, restoring peace to the SETI community.

## 5 CONCLUSIONS

The current speculations are meant to illustrate a crucial point, namely, that an efficient transgalactic communication network should consist not only of permanently placed CPs, but also of flyby scout CPs. It is plausible that the transgalactic communication architecture described in this paper emerged from an earlier architecture wherein individual civilizations sent probes out in concentric spheres, which then, upon encountering one another, linked together to form the lattice postulated here.

SETI is a proto-science, not unlike the alchemy of the Middle Ages. In that era, there were real laboratories in which real experiments with real chemicals, minerals and metals were conducted, resulting in some real understandings of how matter could be manipulated. However, in the absence of any atomic or molecular theory or a periodic table, the alchemists were grasping at straws, and prone to often filling the void of their knowledge with wild flights of imagination in which an astrological influence or a Philosopher's Stone could be invoked to explain anything.

With little more than a pixel to go upon, Loeb probably mischaracterized a very interesting rock as an ET solar sail, just as a low resolution photo of a small Martian mountain taken by Viking led some researchers to wrongly conclude that it was an artificially crafted humanoid face [19]. Indeed, once the true nature and prevalence of technological life in the universe is finally understood, most of our current speculations will probably look as naive as medieval alchemy. Nonetheless, even as alchemists paved the way for a scientific understanding of matter, so too do our current speculations serve as important way stations in the quest for knowledge. The grand experiment, known as SETI, should persevere, employing search strategies informed by these speculations.

### Acknowledgement

The author gratefully acknowledges the kind assistance of Geoff Marcy in the organization of the ideas expressed in this paper.